\documentclass[conference]{IEEEtran}
\usepackage{cite}
\usepackage{amsmath,amssymb,amsfonts}
\usepackage{hyperref}
\usepackage{amsfonts}
\usepackage{graphicx}
\usepackage{subfig}
\usepackage{amssymb,mathtools}
\usepackage{algpseudocode}
\usepackage[linesnumbered,ruled,vlined]{algorithm2e}
\usepackage{soul}
\usepackage{textcomp}
\usepackage{xcolor}

\usepackage[a4paper, total={184mm,239mm}]{geometry}
\def\BibTeX{{\rm B\kern-.05em{\sc i\kern-.025em b}\kern-.08em
T\kern-.1667em\lower.7ex\hbox{E}\kern-.125em}}

\begin{document}

\newif\ifproofread
\newcommand{\changemarker}[1]{%
\ifproofread
\textcolor{blue}{#1}%
\else
#1%
\fi
}

\renewcommand{\IEEEbibitemsep}{0pt plus 0.5pt}
\makeatletter
\IEEEtriggercmd{\reset@font\normalfont\fontsize{7.4pt}{8.0pt}\selectfont}
\makeatother
\IEEEtriggeratref{1}

\title{Uncertainty-Aware Hardware Trojan Detection Using Multimodal Deep Learning}

\author{
\IEEEauthorblockN{Rahul Vishwakarma}
\IEEEauthorblockA{\textit{Computer Engineering \& Computer Science Department} \\
\textit{California State University Long Beach}\\
Long Beach, CA, USA \\
rahuldeo.vishwakarma01@student.csulb.edu} 
\and
\IEEEauthorblockN{Amin Rezaei}
\IEEEauthorblockA{\textit{Computer Engineering \& Computer Science Department} \\
\textit{California State University Long Beach}\\
Long Beach, CA, USA \\
amin.rezaei@csulb.edu} 
}

\maketitle

\begin{abstract}
The risk of hardware Trojans being inserted at various stages of chip production has increased in a zero-trust fabless era. To counter this, various machine learning solutions have been developed for the detection of hardware Trojans. While most of the focus has been on either a statistical or deep learning approach, the limited number of Trojan-infected benchmarks affects the detection accuracy and restricts the possibility of detecting zero-day Trojans. To close the gap, we first employ generative adversarial networks to amplify our data in two alternative representation modalities: a graph and a tabular, which ensure a representative distribution of the dataset. Further, we propose a multimodal deep learning approach to detect hardware Trojans and evaluate the results from both early fusion and late fusion strategies. We also estimate the uncertainty quantification metrics of each prediction for risk-aware decision-making. The results not only validate the effectiveness of our suggested hardware Trojan detection technique but also pave the way for future studies utilizing multimodality and uncertainty quantification to tackle other hardware security problems.
\end{abstract}

\begin{IEEEkeywords}
Hardware Trojan; Multimodal Deep Learning; Uncertainty Quantification
\end{IEEEkeywords}

\section{Introduction}
Hardware Trojan (HT) insertion has become a major concern in today's fabless semiconductor manufacturing since attackers can make malicious modifications for a variety of reasons, such as information leakage, incorrect operation, or inflicting damage on the chip \cite{salmani2017hardware,  Regazzoni:HTDetection, Guin:HTDetection, Salmani:HTDetection}. 
While comprehensive approaches are vital for countering HTs, they entail certain drawbacks. Formal methods and simulation-based testing can be resource-intensive and time-consuming. Intrusion detection systems may yield false alarms, disrupting operations. 
Establishing a secure supply chain can limit flexibility in supplier selection. 

Recently, Machine Learning (ML) has emerged as a powerful tool for detecting HTs \cite{gubbi2023hardware, huang2020survey, liakos2019machine, koblah2023survey, koylu2023survey}. 
It leverages algorithms to identify intricate patterns indicative of Trojans, even in increasingly sophisticated attacks. By training on diverse datasets, ML models can classify circuits as Trojan-free or Trojan-infected. Real-time processing enables continuous monitoring and immediate threat response. However, challenges exist, for example, acquiring large and diverse datasets, especially for rare Trojans, which can be difficult. Moreover, models are susceptible to adversarial attacks \cite{west2023towards}, potentially undermining their decision-making. Interpretability \cite{li2022interpretable} and explainability \cite{caruana2020intelligible} are crucial for trust but can be complex in this context. Additionally, resource-intensive training and deployment may limit accessibility for smaller manufacturers. Continuous retraining is necessary to adapt to evolving Trojan techniques \cite{Vishwakarma:ICCAD}, adding complexity to maintenance. 

Our goal in this paper is to address the gaps in the current ML-based approaches for identification of HTs by 
proposing \textbf{NOODLE}, an u\textbf{N}certainty-aware hardware Tr\textbf{O}jan detecti\textbf{O}n using multimo\textbf{D}al deep \textbf{LE}arning. The proposed method uses graph representation and tabular data and performs binary classification.

\subsection{Related Works}
The emphasis in traditional ML approaches for HT detection has primarily been on modeling techniques. This entails the development and implementation of algorithms aimed at enhancing the overall accuracy of HT detection. 
Many research papers have concentrated on extracting features from Register Transfer Level (RTL) or gate-level netlists and employing ML models such as Support Vector Machine (SVM) \cite{bao2014application}, Neural Network (NN) \cite{hasegawa2017hardware}, eXtreme Gradient Boosting (XGB) \cite{dong2019machine}, and the Random Forest (RF) classifier \cite{hasegawa2017trojan}. In \cite{ashok2022hardware}, image classification techniques are also employed. 

Multimodal Deep Learning (DL) has been a well-explored topic in the Artificial Intelligence (AI) community. Early research, exemplified by Deep Boltzmann Machines (DBM) focused on the model's capacity to understand probability distributions across inputs with multiple modes \cite{srivastava2012multimodal}. Additionally, applications of uncertainty-aware multimodal learning \cite{wang2022uncertaintyaware} have been successfully demonstrated in healthcare \cite{sarawgi2021uncertainty} and in scenarios involving multimodal task distributions \cite{almecija2022uncertaintyaware}, particularly in safety-critical environments. In our work, we target the fusion of graph \cite{ektefaie2023multimodal, kim2023heterogeneous} and Euclidean data as the modalities of interest along with uncertainty estimation. 

Moreover, when working in the hardware security domain, it is expected to have fewer data points that are malicious or Trojan-infected. In this context, it is necessary to work with small data \cite{nyiri2023can} and this has been achieved in various domains such as material science \cite{xu2023small} and anomaly detection \cite{ghamisi2023anomaly}.

\subsection{Contributions}
In this paper, we investigate the feasibility of applying a multimodal ML approach for HT identification by deriving two data representations of circuits. The first is graphical representation \cite{yu2021hw2vec} of circuits, and the second is euclidean data \cite{px6s-sm21-22} derived by processing the Abstract Syntax Tree (AST) of the RTL files (Verilog). Although the use of multimodal approaches for improved model accuracy has been used in other domains, we do not see any application in Trojan identification. For uncertainty-aware multimodal learning, we believe the logic should be implemented at the information fusion level of the modalities, and thus, we leverage the $p$-values aggregation with conformal prediction. Our main contributions are as follows:
\begin{itemize}
    \item Proposing a multimodal learning approach using graph and euclidean data of the hardware circuits. To the best of our knowledge, this study is the first to investigate and implement a multimodal approach for HT detection.
    \item Suggesting a model fusion approach using $p$-values with an uncertainty quantifier. By employing $p$-values as statistical measures, we can systematically assess each modality's contribution to the overall prediction. This not only enhances the interpretability of the fusion process but also enables more robust decision-making.
    \item Addressing the challenges of missing modalities and solving the issue of handling an imbalanced and small dataset by leveraging generative adversarial networks.
\end{itemize}

\begin{figure*}[ht]
  \centering
   \includegraphics[width=1\textwidth]{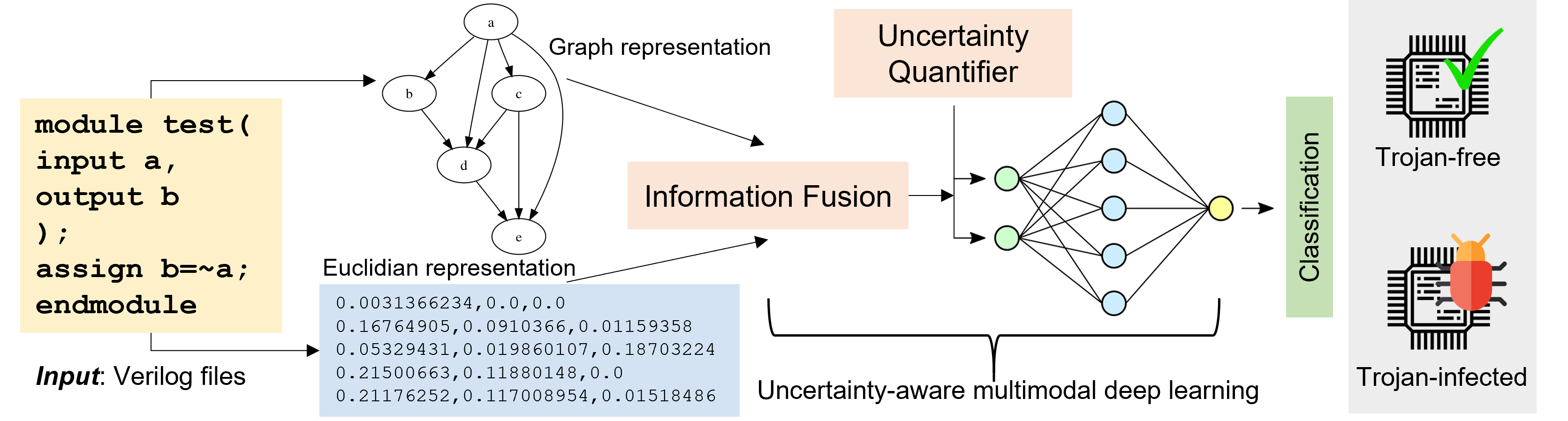}
   \caption{NOODLE framework: The input consists of an RTL file (Verilog), which undergoes conversion into both graph and Euclidean representations, and then input into a multimodal deep learning classifier. This classifier yields a decision indicating whether the circuit is Trojan-infected or Trojan-free.}
  \label{fig:intro}
\end{figure*}

\section{Preliminaries}
\label{Sec:Prelem}

\subsection{Multimodal Learning}
\label{sec:multimodal}
Multimodal learning \cite{ngiam2011multimodal} addresses complex problems by integrating information 
from multiple modalities, such as text, images, and audio, to obtain a comprehensive understanding of a given phenomenon. In our case, we use graphical data and tabular representations of the source circuits. This approach enables models to capture nuanced relationships that may be overlooked when considering each modality in isolation, and thus empowers the model to make more robust predictions.

From a mathematical perspective, multimodal learning involves the integration of data representations into a unified framework. Let \(X_1, X_2, ..., X_M\) represent \(M\) different modalities of data, each with their respective feature spaces \(\mathcal{F}_1, \mathcal{F}_2, ..., \mathcal{F}_M\). The task is to learn a mapping \(f\) that captures the relationships between these modalities. Mathematically, this can be formulated as:
\begin{equation}
f: \mathcal{F}_1 \times \mathcal{F}_2 \times ... \times \mathcal{F}_M \rightarrow \mathcal{Y}
\end{equation}
where \(\mathcal{Y}\) is the target space, representing the desired prediction.

The challenge lies in effectively combining information from diverse modalities, which can be approached through various techniques such as late fusion or early fusion. 

In late fusion \cite{trong2020late}, features are extracted independently from each modality and then combined at a later stage. This approach treats modalities as separate entities until a decision needs to be made and can be represented as:
\begin{equation}
f(x_1, x_2, ..., x_M) = g(h_1(x_1), h_2(x_2), ..., h_M(x_M))
\end{equation}
where \(h_i\) represents feature extraction for modality \(i\), and \(g\) combines the extracted features.

In early fusion \cite{nguyen2021gefa}, information from different modalities is combined at the input level, resulting in a joint feature representation which can be expressed as:
\begin{equation}
f(x_1, x_2, ..., x_M) = h(x_1, x_2, ..., x_M)
\end{equation}
where \(h\) combines the raw input data from all modalities.


\subsection{Calibrated Prediction}
\label{sec:Calibration}
Calibration involves ensuring that a model's confidence score accurately reflects the true probability of the prediction's correctness \cite{ovadia2019can}. Let $X$ be the input data, and $Y$ be the output label. Given a training dataset $D = {(x_1, y_1), (x_2, y_2),..., (x_n, y_n)}$, the goal is to learn a function $f$ that can predict the correct output label $y$ for a given input $x$. The output of the model for an input $x$ can be denoted as $f(x)$, and the true probability of the prediction's correctness can be denoted as $P(y=1|x)$. A calibrated model produces a confidence score $g(x)$ that reflects the true probability of correctness of the prediction. The goal of calibration is to ensure that the confidence score $g(x)$ is well-calibrated, i.e., $P(y=1|g(x)=p) = p$ for all $p$ in the range $[0, 1]$.

Calibration is a crucial aspect in HT detection since it aids in determining the likelihood of the existence of a Trojan in a circuit, which can have a significant impact on decision-making. In situations where a model's confidence score is high, but the likelihood of a Trojan's presence is low, it is reasonable to assume that the circuit does not contain a Trojan. Conversely, if the confidence score is low but the likelihood of a Trojan's presence is high, further investigation of the circuit is necessary.

\begin{algorithm}[!b]
\small
\SetKwInOut{Input}{Input}
\SetKwInOut{Output}{Output}
\Input {Number of data sources $N$; \newline
Training sets for each data source $T_1 = \{(x^{(1)}_1, y_1), \ldots, (x^{(1)}_n, y_n)\}, \ldots, T_N = \{(x^{(N)}_1, y_1), \ldots, (x^{(N)}_n, y_n)\}$, where $x^{(j)}_i$ is the $i$th data point belonging to the $j$th data source and $y_i$ is the class label of the $i$th data point; \newline
Number of classes $M$; \newline
Class labels $y^{(i)} \in Y = \{y^{(1)}, y^{(2)}, \ldots, y^{(M)}\}$; \newline
Classifiers $S_1, \ldots, S_N$ for each data source; \newline
Confidence level $E$.}

\Output {Conformal prediction regions $r_E = \{y^{(j)} : \hat{p}_j > 1 - E, y^{(j)} \in Y\}$.}

Get the new unlabeled example w.r.t each data source $x^{(1)}_{n+1}, \ldots, x^{(N)}_{n+1}$.

Evaluate conformal predictors and classifiers $S_1, \ldots, S_N$ corresponding to each data source, compute $p$-values $p^{(i)}_j$, where $i = 1, \ldots, N$ corresponds to the $i$th data source and $j = 1, \ldots, M$ corresponds to the $j$th class label.

\For {each class label $y^{(j)}$, $j = 1, \ldots, M$}{
    Compute $p$-value, $\hat{p}_j$, of combined hypothesis from $N$ modalities}

\Return $r_E$.
\caption{Uncertainty-aware information fusion}
\label{algo:mcp}
\end{algorithm}

\subsection{Conformal Prediction}
\label{sec:CP}
Conformal Prediction (CP) is a ML framework that assesses prediction uncertainty by generating sets of possible predictions \cite{shafer2008tutorial}. This approach strengthens the inference of conventional models, ensuring their reliability and enabling confidence estimation for individual predictions. It is worth noting that minority classes often bear a disproportionate burden of errors when label-conditional validity is lacking \cite{lofstrom2015bias}. Nevertheless, this challenge can be mitigated by ensuring label-conditional validity, which guarantees that the error rate, even for the minority class, will eventually converge to the selected significance level in the long run. 

In certain instances, CP may yield uncertain predictions, signifying that prediction sets contain more than one possible value. This happens when none of the labels can be rejected at the specified significance level. Moreover, when employing CP, the confusion matrix differs from the conventional one due to the distinctive nature of prediction sets, which encompass multiple values rather than a single one. 
In cases where providing a single-point prediction may be more appropriate than a prediction set or interval in a hedged forecast, opting for the label with the highest $p$-value is a straightforward and reasonable choice. 

There has been limited exploration of the application of CP in modal fusion \cite{balasubramanian2015conformal}. This method entails treating individual data sources as separate hypothesis tests within the CP framework. Subsequently, it utilizes $p$-value combination techniques as a test statistic for the combined hypothesis after the fusion process. Our approach relies on the Mondrian Inductive Conformal Prediction (ICP) methodology \cite{bostrom2021mondrian} outlined in Algorithm \ref{algo:mcp} for the uncertainty-aware fusion of various modalities during classification. This algorithm can be effectively extended for both early and late fusion of the modalities.

 


\begin{algorithm}[!b]
\small
\SetKwInOut{Input}{Input}
\SetKwInOut{Output}{Output}
\Input {RTL-level files (Verliog) of circuits}
\Output {Decision (D) = Trojan-free or Trojan-infected}
\For {each circuit $C$}{
    Convert $C$ to Graph data \textbf{G} and Euclidean data \textbf{T}.
    \newline 
    \If{$\exists$ missing modalities}{
       perform GAN to impute the missing modality. 
    }
}
Feed the modalities to \textit{CNN}-based classifier.
\newline 
\For {each modalities $M$}{
    Use Algorithm \ref{algo:mcp} for uncertainty-aware information fusion.\\
    Perform early fusion.\\
    Perform late fusion. 
}
Choosing the winning fusion method.\\
\Return $D$.
\caption{Multimodal deep learning}
\label{algo:mdd}
\end{algorithm}

\section{Multimodal Hardware Trojan
Detection}
\label{sec:solution}
While state-of-the-art works on HT detection have focused mainly on choosing the right algorithm and choosing different representations of the dataset for improved accuracy, incorporating different modalities of the same data and feeding it to the ML system has not been investigated. By performing information fusion derived from different modalities, a more refined data representation can be achieved. Furthermore, in a practical scenario, we encounter missing values while collecting data, and this may lead to missing modalities when dealing with a multimodal ML approach. So, we also need a method that handles missing modalities for any given dataset. Lastly, in the domain of hardware security, it is difficult to get enough data for training, especially the labels marked as Trojan-infected because of the rarity of the event. In such a situation, we need to work with limited data.


Our proposed \textit{NOODLE} framework is shown in Fig. \ref{fig:intro} emphasizing the design and implementation, and a pseudocode is also provided in Algorithm \ref{algo:mdd}. We choose to use two modalities, i.e., graph and tabular data representations. Methods like multimodal autoencoder \cite{jaques2017multimodal} have been used for handing missing modalities; however, we use Generative Adversarial Networks (GANs) \cite{creswell2018generative} to increase the dataset size to 500 data points as it aims to generate samples that are consistent with the joint distribution of the observed modalities and facilitate more effective multimodal fusion. The data points labeled as Trojan-Free (TF) will be segregated, and only these will be used to generate more data points using GAN so that they represent the same label, and we will do the same for data labeled as Trojan-Infected (TI). Before performing multimodal learning, we first explain the working of uncertainty-aware model fusion.

To perform an uncertainty-aware multimodal fusion, we leverage conformal prediction $p$-values for the model fusion as described in Algorithm \ref{algo:mcp}. First, we use a Convolutional Neural Network (CNN)-based classifier for graph and tabular data sources with a designed non-conformity score that provides $p$-values for each label and for each data modal. The below non-conformity score can be used in the CP framework to get calibrated conformal predictions:
\begin{equation}
NS = \sum_{t=1}^T B_t(x, y)
\end{equation}
where \(B_t(x, y)\) is the non-conformity score of \((x, y)\) computed from a classifier, \(h_t\). Thus, for every class label \(y(j)\), \(j \in \{1, ..., M\}\), we have an individual null hypothesis for each data source, \(H0_1, H0_2, ..., H0_N\), where \(M\) is the number of class labels, which in our case is either TF or TI, and \(N\) is the number of data sources. Thus, for every class label \(y(j)\), we obtain \(N\) $p$-values, \(p(i)\), \(i = 1,..., N\) (one for each modality). These $p$-values are then combined into a new test statistic \(C(p(1), ..., p(N))\), which is used to test the combined null hypothesis \(H0\) for class label \(y(j)\). 
The conformal prediction region at a specified confidence level, \(r_E\), is then presented as a set containing all the class labels with a $p$-value greater than \(1-E\). The mentioned steps helps in realization of uncertainty-aware multimodal fusion.

After obtaining a sufficient number of data points for the experiment, we implement multimodal ML using the graph and tabular data. Specifically, we have employed a CNN for binary classification. It is worth mentioning that any ML model can be optimized through hyperparameter tuning to enhance accuracy. However, our primary emphasis is on assessing the effectiveness of uncertainty-aware multimodality by accessing early and late fusions. Finally, the model will be used to make further informed decisions for the detection of HTs.

\section{Experimental results}
\label{sec:Results}
We used Python (3.9) and implemented \textit{NOODLE} on macOS (13.3.1) with 8GB RAM. The experimental results with source code and the dataset are hosted on GitHub\footnote{https://github.com/cars-lab-repo/NOODLE}.

\subsection{Dataset}
For our experiment, we have used the features extracted from the TrustHub RTL-level (Verilog) Trojan dataset based on code branching features \cite{px6s-sm21-22} and the graph dataset in \cite{yu2021hw2vec} which includes RTL source code files (Verilog) for each IP core design containing both malicious and non-malicious functions.

{\renewcommand{\arraystretch}{1.2}%
\begin{table}[ht]
\caption{Brier score comparison for different modalities}
\centering
\begin{tabular}{lc}
\hline
\textbf{Dataset} & \textbf{Brier Score} \\ \hline \hline
Graph-based Data & 0.1798 \\ \hline
Tabular-based Data & 0.1913 \\ \hline
NOODLE - Early Fusion (Graph + Tabular) & 0.1685 \\ \hline
NOODLE - Late Fusion (Graph + Tabular)  & 0.1589 \\ \hline
\end{tabular}
\label{tab:table1}
\end{table}
}

\subsection{Brier Score}
For any of the classification problem statements, the most common performance metric is model accuracy, followed by various other complementing metrics such as precision recall and F1-score. However, these metrics can be misleading in situations where the class distribution is imbalanced, as in our case. For this reason, we have used the Brier score as an evaluation metric for assessing the quality of probabilistic predictions in the classification of HTs. The Brier score, which
offers insights into accuracy and calibration, is defined as follows:
\begin{equation}
BS = \frac{1}{N} \sum_{i=1}^{N} (p_i - o_i)^2
\end{equation}
where \(N\) is the number of instances, \(p_i\) is predicted probability for instance \(i\), and \(o_i\) is the observed outcome for instance \(i\). The Brier score ranges from 0 to 1. A score of 0 indicates perfect accuracy, meaning the predicted probabilities perfectly match the actual outcomes. A score of 1 signifies complete inaccuracy, where the predicted probabilities are entirely different from the actual outcomes.

\begin{figure}[!t]
\centering
  \subfloat[]
  {\includegraphics[width=0.5\columnwidth]{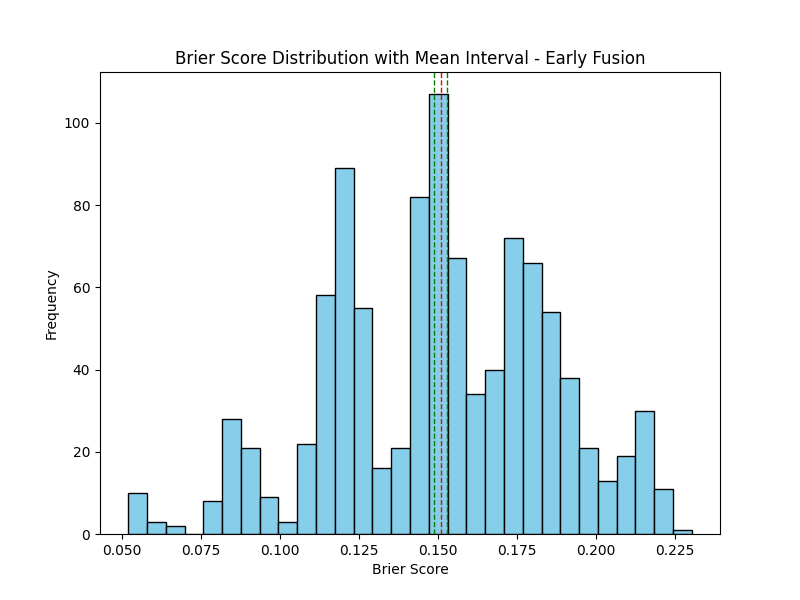}}
  \subfloat[] 
  {\includegraphics[width=0.5\columnwidth]{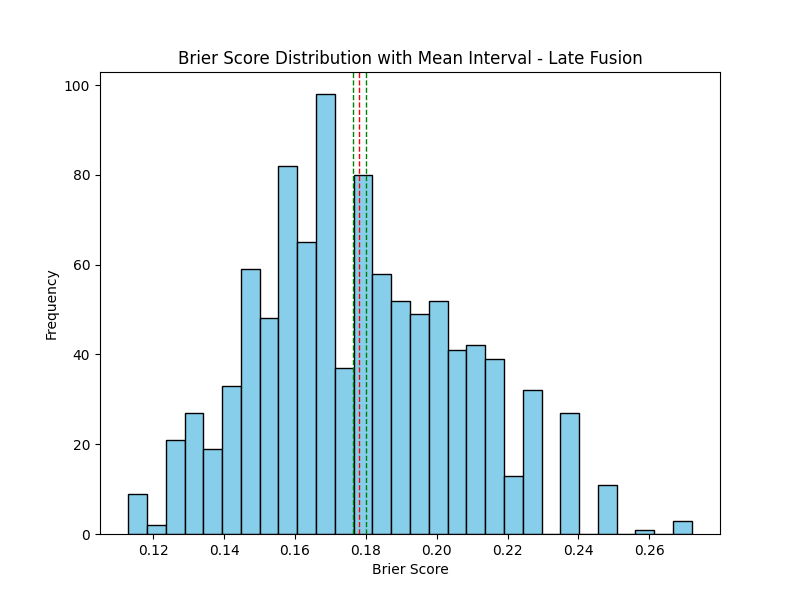}}
  \caption{NOODLE's Brier score (a) Early fusion (b) Late fusion}
  \label{fig:gan}
\end{figure}

\begin{figure*}
\centering
\begin{minipage}{0.30\textwidth}
\centering
\includegraphics[width=0.79\textwidth] {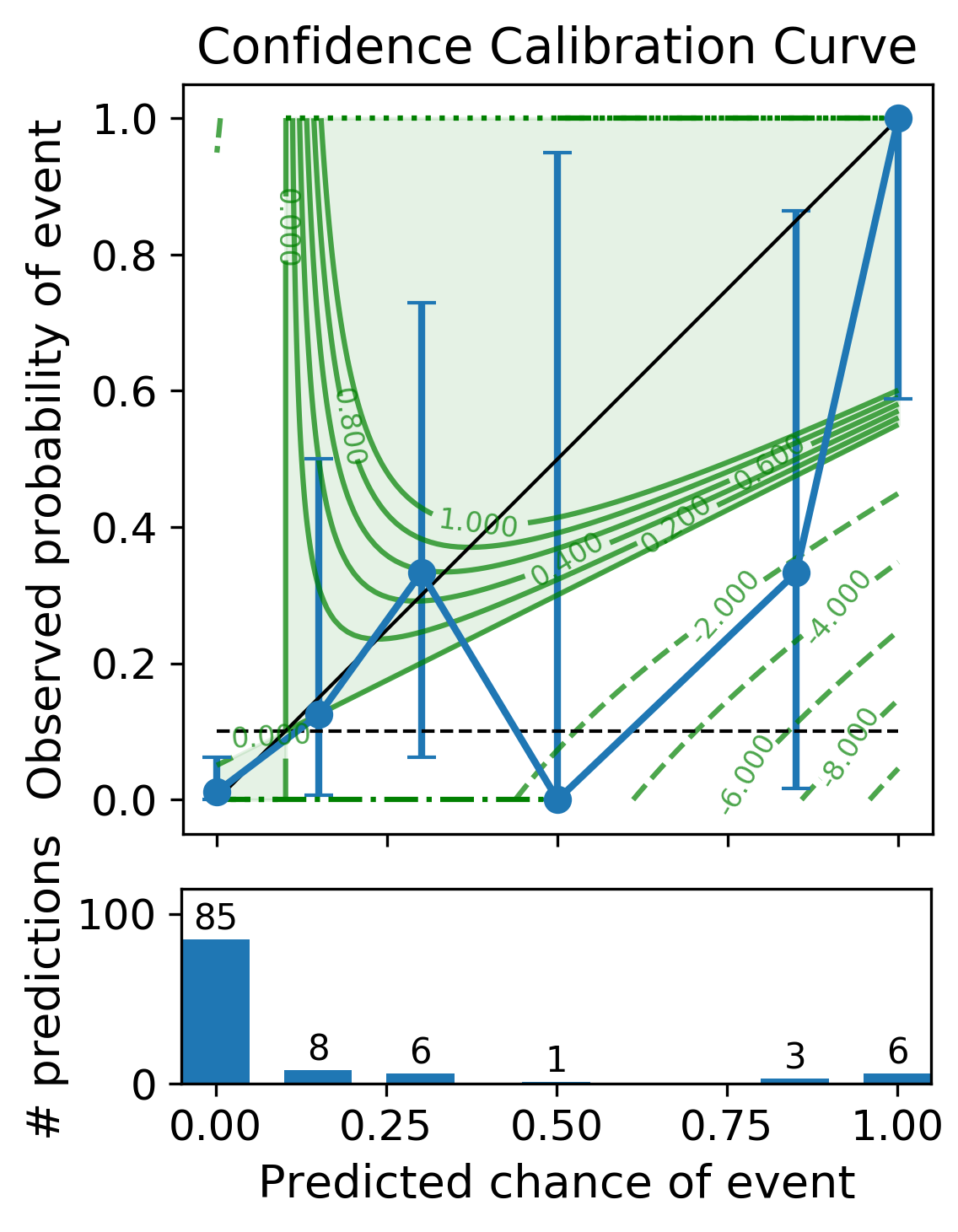}
\captionof{figure}{NOODLE's confidence calibration curve}
\label{fig:ccc}
\end{minipage}%
\hspace{0.5em}
\begin{minipage}{0.30\textwidth}
\centering
\includegraphics[width=0.98\textwidth]{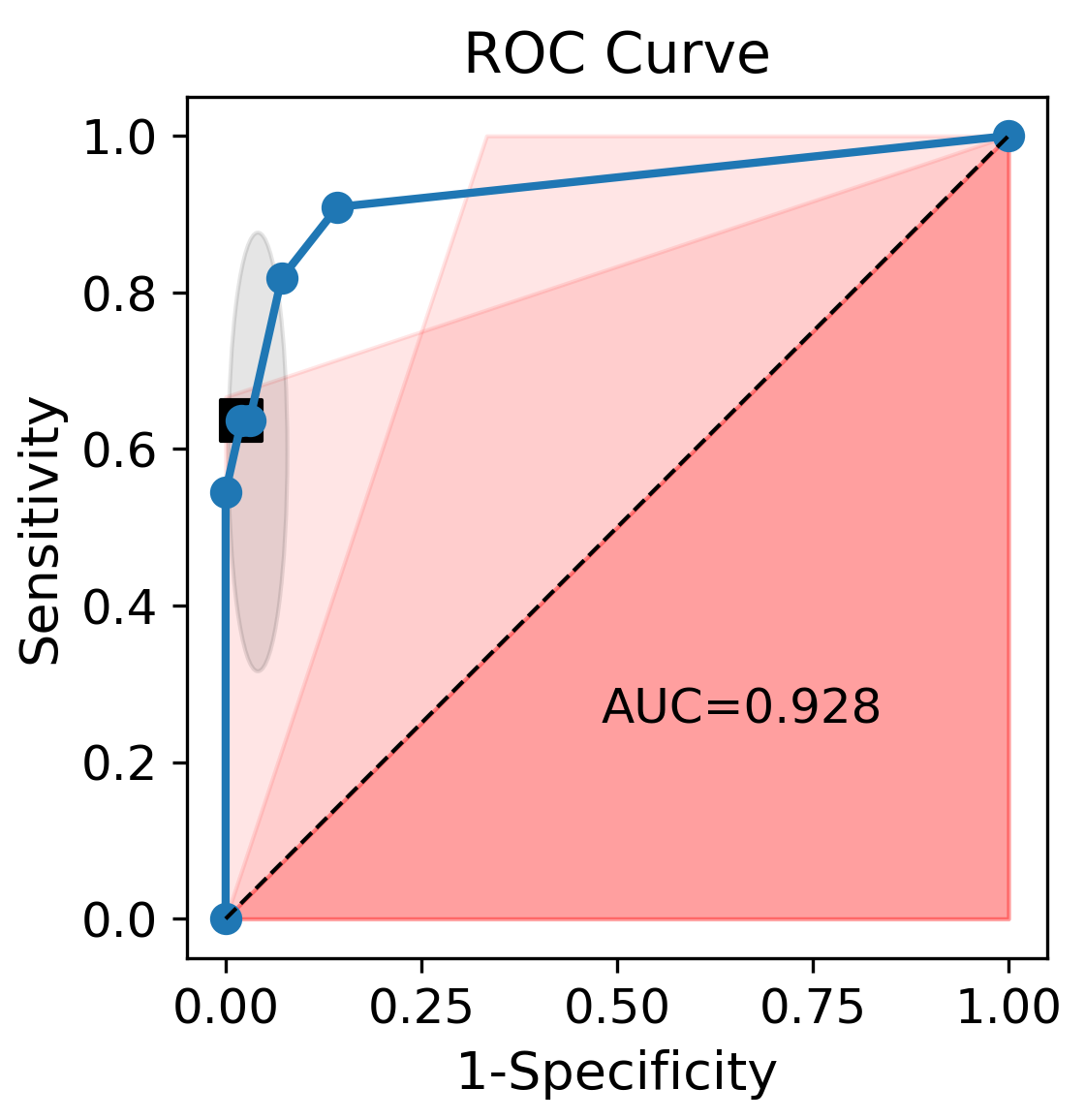}
\captionof{figure}{NOODLE's ROC-AUC curve under late fusion}
\label{fig:gan5}
\end{minipage}
\hspace{0.5em}
\begin{minipage}{0.35\textwidth}
\centering
\includegraphics[width=\textwidth]{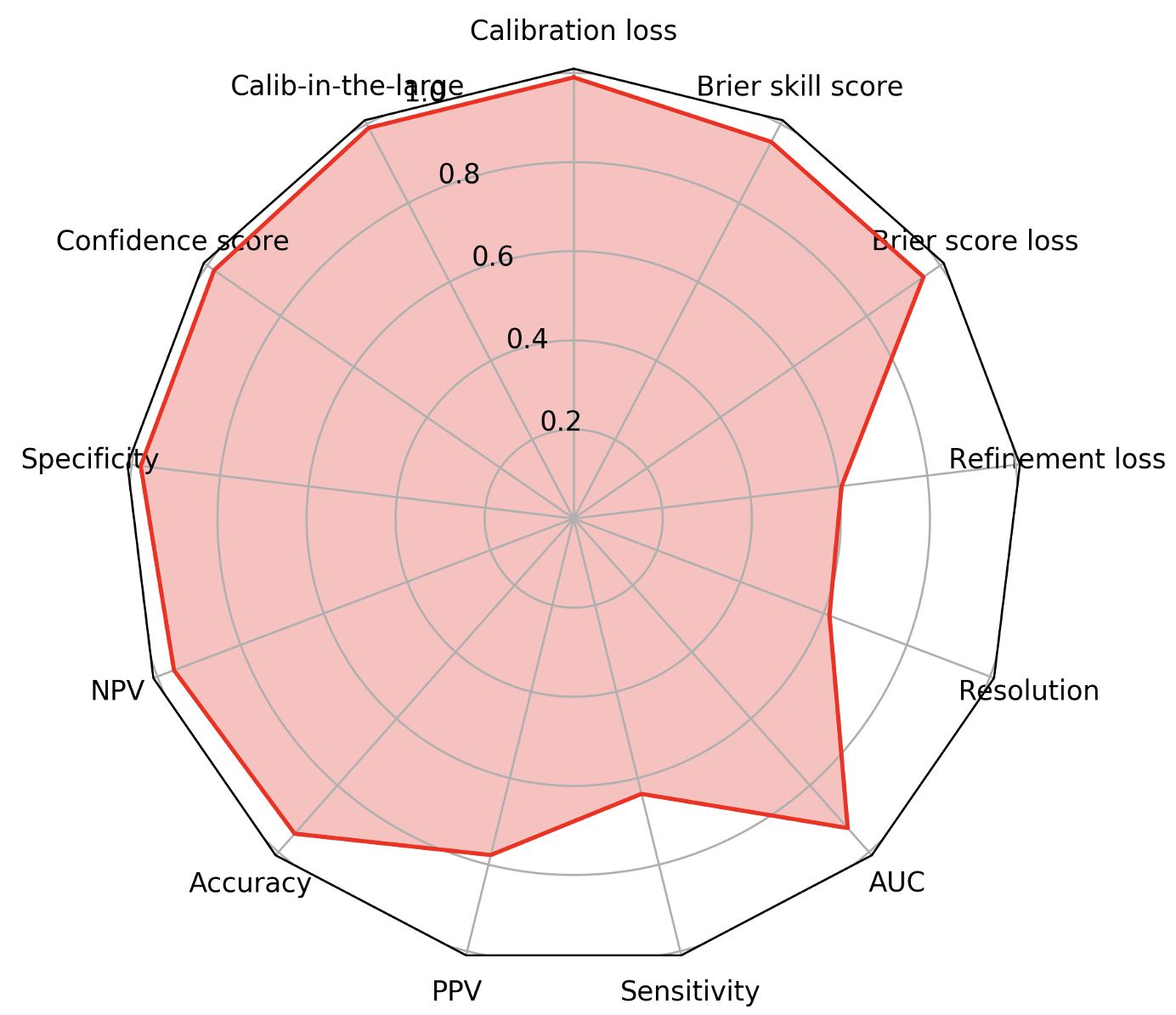}
\captionof{figure}{NOODLE's radar plot for consolidated metrics}
\label{fig:radar}
\end{minipage}
\end{figure*}


We begin the evaluation process by independently assessing each modality. This involves conducting binary classification on both the graph dataset and the tabular data. The resulting comparative Brier scores for these classification tasks are presented in Table \ref{tab:table1}. The experimental outcome demonstrates that, when employing the same CNN-based deep learning model with identical hyperparameters, the graph dataset yields a superior Brier score (0.1798) compared to the tabular data (0.1913). It is worth noting that while we established a baseline model using CNN, any other alternative classification algorithms can also be employed in this context.

Then, we tested \textit{NOODLE} with two different information fusion approaches, i.e., early fusion (feature) and late fusion (decision). As shown in Table \ref{tab:table1}, the early fusion approach, which combines the graph and tabular data before processing, yields a Brier score of 0.1685. On the other hand, the late fusion strategy, which integrates the graph and table data after individual processing, demonstrated the best performance with a Brier score of 0.1589.

It is worth noting that neither of these data fusion methods can be deterministically labeled as superior \cite{gallo2017multimodal} as each one of them will demonstrate their potential to produce favorable outcomes when the data distribution changes. For this reason, we implemented both of the fusion approaches and chose the approach that provides a better Brier score (i.e., closer to 0), as mentioned in Step 8 of Algorithm \ref{algo:mdd}. The corresponding Brier score distribution with mean interval is also shown in Fig. \ref{fig:gan}a and Fig. \ref{fig:gan}b for early and late fusion, respectively. This provides a comprehensive view of predictive accuracy across multiple scenarios and is also useful for comparing models and understanding the variability in performance.



\subsection{Confidence Calibration Curve}
The confidence calibration curve 
plots observed probabilities of occurrence as a function of the predicted probabilities for the classification model, as shown in Fig. \ref{fig:ccc}. For the model to be perfectly calibrated, it will have all data points along the diagonal; however, in our case, the model is not well calibrated because of the highly imbalanced dataset. These are the cases on which any decision-maker should focus while making a risk-aware decision and not completely relying on accuracy alone. It helps evaluate the alignment between a model's predicted probabilities and the actual likelihood of events.

A histogram at the bottom of Fig. \ref{fig:ccc} shows the predicted chance for 109 test data. It describes the distribution of the forecasts and helps with visualization of the sharpness, i.e., tendency of the predictions to lie at the extremes of the 0-1 distribution, and is equal to the variance of the predictions.


\subsection{ROC-AUC Curve}
The Receiver Operating Characteristic (ROC) curve illustrates the balance between sensitivity and specificity in a model. It provides a visual representation of how these two metrics change as the threshold for classifying a condition varies. The Area Under the Curve (AUC), on the other hand, quantifies the likelihood that a randomly chosen pair of circuits, one with the Trojan and one without, will be accurately classified by the model. The \textit{NOODLE}'s ROC-AUC curve is given in Fig. \ref{fig:gan5}.

The white area represents the optimal zone for model performance, and the lightly shaded red areas represent the zones of acceptable efficacy. The values for ROC-AUC range from 0 to 1, where values near `1' suggest that it can effectively discriminate between TF and TI cases with a high degree of confidence, and if the value is near `0', the model's performance is worse than random guessing. In our case, the value is 0.928, which suggests that the model is performing well.

\subsection{Radar Plot}
The radar plot provides a visual means of presenting complex, multi-dimensional data, as shown in Fig. \ref{fig:radar}. When appraising the effectiveness of a predictor, there is a tendency to focus narrowly on a limited set of metrics. However, the radar plot provides a method for gaining a comprehensive understanding of performance across diverse dimensions. In a radar chart, each variable is represented along its corresponding axes (some variables have been normalized to conform to the 0-1 range of the radial axis). It is also important to organize the variables in a way that clusters connected ideas or principles. This aids in conducting a thorough evaluation of various facets of performance.

In the given radar plot, we have metrics related to discrimination, which include AUC, resolution, and refinement loss. Following these are combined metrics assessing both calibration and discrimination, namely the Brier score and Brier skill score. As shown in the figure, the model is less sensitive and has high accuracy. This implies that while the model is generally accurate in its predictions, it may not be as effective in identifying all the actual TI cases. This could be due to a higher number of false negatives, which means the model is missing some of the positive cases.


\section{Conclusion}
\label{Sec:Conclusion}
In this paper, we have addressed the growing concern of maliciously inserted hardware Trojans into chips at various stages of production in an era where fabless manufacturing is hard to trust. Specifically, we adopted an innovative approach by utilizing generative adversarial networks to expand our dataset with two distinct representation modalities: graph and tabular. Additionally, we introduced an uncertainty-aware multimodal deep learning framework called \textit{NOODLE} for detecting hardware Trojans. We assessed our findings using both early and late fusion strategies, offering a comprehensive evaluation of our approach's efficacy. Moreover, we integrated metrics for uncertainty quantification for each prediction, enabling us to make decisions that are mindful of potential risks. The utilization of multimodality and uncertainty quantification shows great potential for addressing other critical challenges in hardware security such as logic locking \cite{Rezaei:BreakUnroll, Rezaei:PUF, Maynard:DK-Lock, Aghamohammadi:CoLA}. These contributions collectively represent a significant step forward in enhancing the security and reliability of hardware systems in the face of emerging threats.

\section*{Acknowledgment}
This material is based upon work supported by the National Science Foundation under Award No. 2245247.

\bibliographystyle{IEEEtran}
\bibliography{IEEE}

\end{document}